\documentclass[twocol]{ametsocV6.1}

\usepackage{multirow}
\usepackage{booktabs}
\usepackage{subcaption}
 
\title{Physics-constrained deep learning postprocessing of temperature and humidity}

\authors{
    Francesco Zanetta\aff{a}\aff{b}\correspondingauthor{Francesco Zanetta, zanettaf@ethz.ch}Daniele Nerini\aff{b} Tom Beucler\aff{c} Mark A. Liniger\aff{b}
}

\affiliation{
    \aff{a}{Institute for Atmospheric and Climate Science, ETH Zürich, Zürich, Switzerland}\\
    \aff{b}{Federal Office of Meteorology and Climatology MeteoSwiss, Locarno, Switzerland}\\
    \aff{c}{Institute of Earth Surface Dynamics, University of Lausanne, Lausanne, Switzerland}
}

\abstract{Weather forecasting centers currently rely on statistical postprocessing methods to minimize forecast error. This improves skill but can lead to predictions that violate physical principles or disregard dependencies between variables, which can be problematic for downstream applications and for the trustworthiness of postprocessing models, especially when they are based on new machine learning approaches. Building on recent advances in physics-informed machine learning, we propose to achieve physical consistency in deep learning-based postprocessing models by integrating meteorological expertise in the form of analytic equations. Applied to the post-processing of surface weather in Switzerland, we find that constraining a neural network to enforce thermodynamic state equations yields physically-consistent predictions of temperature and humidity without compromising performance. Our approach is especially advantageous when data is scarce, and our findings suggest that incorporating domain expertise into postprocessing models allows the optimization of weather forecast information while satisfying application-specific requirements.} 

\begin{document}

\maketitle

%
%
%
\statement
Postprocessing is a widely-used approach to reduce forecast error using statistics, but it may lead to physical inconsistencies. This can be problematic for trustworthiness and downstream applications. We present the first machine learning-based postprocessing method intentionally designed to strictly enforce physical laws. Our framework improves physical consistency without sacrificing performance, and suggests that human expertise can be incorporated into postprocessing models via analytic equations.
%
%
%

%

\section{Introduction}

Weather forecasting centers heavily rely on statistical methods to correct and refine numerical weather prediction (NWP) outputs, which improves skill at low computational cost \citep{hemri_trends_2014, vannitsem_statistical_2021}. While the fundamental approach has remained the same for decades -- statistically relating past NWP model outputs and other additional data, such as topographic descriptors or seasonality, to observations -- the traditional divide between physical and statistical modeling is narrowing as increasingly more sophisticated models emerge to harness the growing volume of available data \citep{vannitsem_statistical_2021}.

Current research focuses particularly on machine learning (ML) techniques, with deep learning (DL) and artificial neural networks (ANNs) emerging as a modern class of post-processing methods with the potential to outperform traditional approaches in several aspects. For example,
\citet{rasp_neural_2018} found that simple feedforward ANNs could significantly outperform traditional regression based post-processing techniques, while being less computationally demanding at inference time. The authors highlighted that ANNs could better incorporate non-linear relationships in a data-driven fashion, and were more suited to handle the increasing volumes of model and observation data thanks to their flexibility. ANNs have also been combined with other statistical techniques such as Bernstein polynomials \citep{bremnes_ensemble_2020} for non-parametric probabilistic predictions.

Furthermore, more sophisticated ANNs such as convolutional neural networks (CNNs) have the ability to incorporate spatial and temporal data with unprecedented flexibility.  \citet{gronquist_deep_2021} used CNNs to improve forecasts of global weather. \citet{hohlein_comparative_2020} and \citet{veldkamp_statistical_2021} used CNNs for spatial downscaling of surface wind fields. A process-specific application was proposed by \citet{chapman_improving_2019, chapman_probabilistic_2022}, with the goal of improving the prediction of atmospheric rivers, which are filaments of intense horizontal water vapor transport \citep{ralph_defining_2018}. \citet{dai_spatially_2021} implemented a generative adversarial network based on CNNs to produce physically realistic post-processed forecasts of cloud cover. Thus, first attempts at using DL-based approaches have shown promising improvements over traditional approaches, as they better capture non-linear dependencies and often require less feature engineering. Still, a number of challenges remain in applying ML approaches to the post-processing world \citep{vannitsem_statistical_2021, haupt_towards_2021}, and they cannot be considered a panacea for all problems. This stresses the need to include more domain expertise into data-driven approaches in a hybrid manner, which is facilitated by the availability of custom losses and architectures in standard machine learning libraries \citep{ebert-uphoff_cira_2021}. 

When traditional post-processing methods are applied, the goal is to minimize the forecast error. This often leads to predictions that do not exhibit the typical spatial and temporal correlation structure that emerges from common patterns of atmospheric phenomena, or predictions that violate physical principles and dependencies between variables. However, for various applications, such as animated maps of meteorological parameters commonly disseminated to the public, or in the context of hydrological forecasting \citep{cloke_ensemble_2009} and renewable energy \citep{pinson_chapter_2018}, it is important to provide forecast scenarios that not only have a smaller error, but also exhibit realistic spatio-temporal structures \citep[e.g.][for related work]{schefzik_ensemble_2017}. Furthermore, consistency across variables should be ensured in various applications. For hydrological modeling, for example, temperature, radiation, and precipitation should be consistent at all times. The issue of consistency is particularly relevant in the context of probabilistic postprocessing, where sampling from marginal predictive distributions is an additional step that further breaks the spatiotemporal and inter-variable consistency. Existing approaches try to model dependencies from a statistical perspective, and not a physical one. We believe the two are complementary, closely related yet different as noted by \cite{mohrlen_chapter_2023}. In the postprocessing field, limited research is available on the issue of physical consistency, in the sense of respecting physical principles or variable dependencies based on analytic relationships. However, this question has recently gained a lot of attention in the wider ML community, and some applications in weather modeling are reviewed in \citet{kashinath_physics-informed_2021} or \citet{willard_integrating_2022}. In general, it has been shown that physical consistency can be pursued by applying constraints to DL models in order to prescribe specific physical processes. These constraints can take many forms. The most widely used approach is to incorporate physics via soft constraints, by defining physics-based losses in addition to common performance metrics such as mean absolute error \citep{daw_physics-guided_2021}. Another popular approach is to design custom model architectures such that the physical constraints are strictly enforced \citep[e.g.][]{beucler_enforcing_2021, ling_reynolds_2016}.

In this paper we explore ways to incorporate domain knowledge in DL-based postprocessing models of temperature and humidity, and the related state variables. Specifically, we evaluate the effect of imposing constraints based on the ideal gas law and an empirical approximation of a thermodynamic state equation, and we identify benefits and disadvantages of different approaches. The goal of this paper is not to develop a highly optimized model for operational use, but rather to provide some technical guidelines and insights about incorporating meteorological expertise, in the form of analytic equations, in postprocessing models of NWP. For this reason, we simplified the problem under several aspects as explained in the next section. Most importantly, we focus here on a deterministic setting, although we hope to extend this framework to probabilistic predictions in the future. 

\section{Data and methods}

\subsection{Datasets}
\begin{figure}[ht]
    \centerline{\includegraphics[width=20pc]{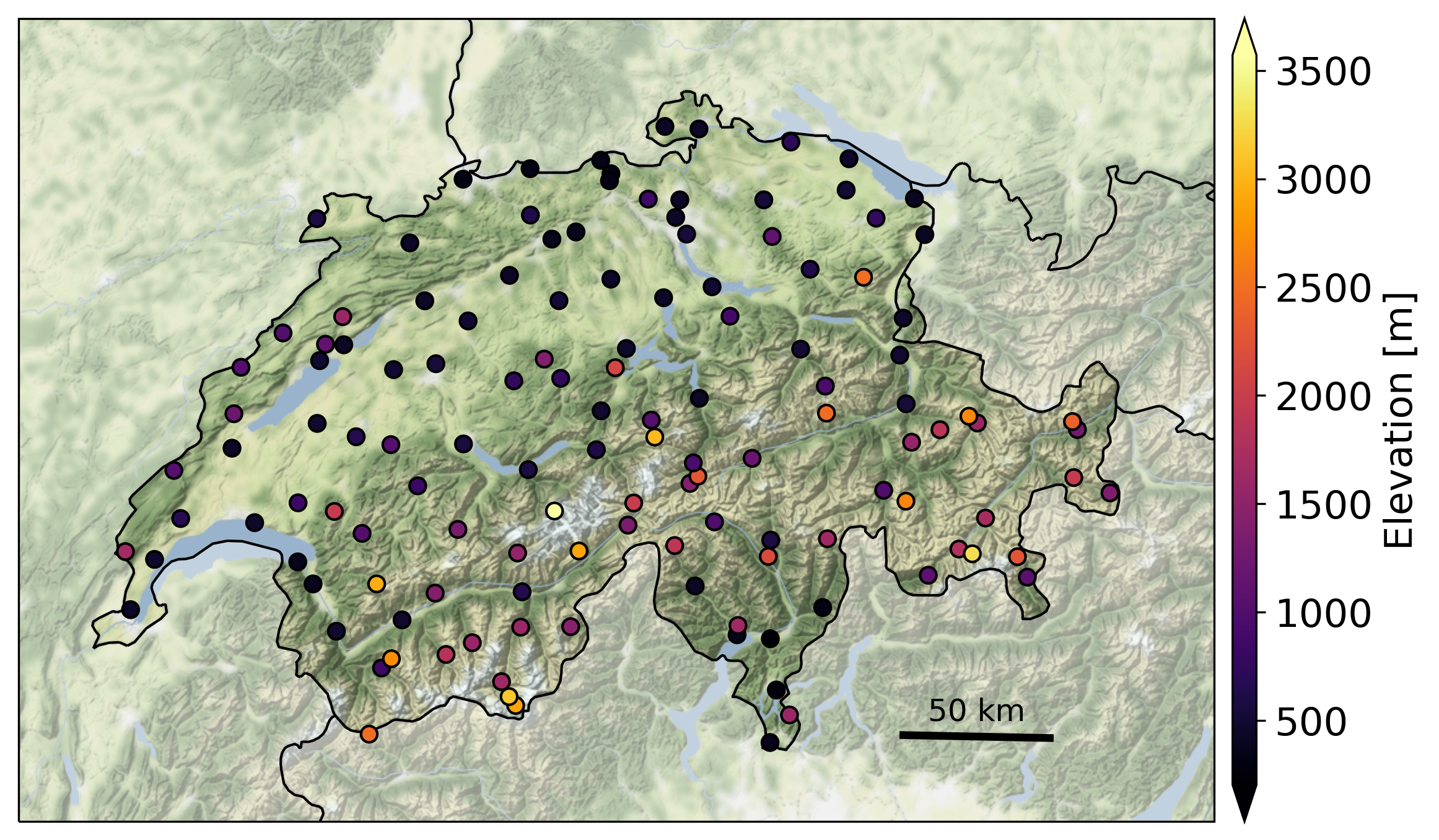}}
    \caption{The location of the 131 weather stations across Switzerland considered in our study. Stations are colored based on their elevation above sea level. We show the topography of Switzerland and its surroundings in the background. Map tiles by Stamen Design, under CC BY 3.0. Data by OpenStreetMap, under ODbL.}
    \label{fig:map}
\end{figure}

In this study, we make predictions at 131 weather station locations covering Switzerland, as shown in Fig.~\ref{fig:map}. We consider forecast data from COSMO-E \citep{klasa_evaluation_2018} for the postprocessing task that spans January 2017-December 2022. COSMO-E is a limited area weather forecasting model used for the operational weather forecasts for Switzerland by MeteoSwiss. It is operated as a 21-member ensemble with a 2.2\.km horizontal resolution. It runs two cycles per day (0000 and 1200 UTC) with a time horizon of five days. We interpolate the nearest model grid-cell to the selected station locations, using the $k$-d tree method. We retain model runs initialized at 0000 UTC and consider leadtimes between +3 and +24 hours with hourly steps. We use observational data from the SwissMetNet network \citep{meteoswiss_automatic_2022} to define our ``ground truth''. The predictors and predictands used in this study are summarized in Table \ref{tab:mapping}. They represent instantaneous measurements with hourly granularity. We note that while using both dew point temperature and dew point deficit is redundant, this is necessary to guarantee a fair comparison between the different model architectures in Sec.~\ref{sec:results}. In contrast to prior postprocessing studies, we train a single model for all leadtimes and use the leadtime as a predictor, thereby increasing the variance in the training set. This choice was motivated by the ease of implementation, allowing us to focus on the main aspect of the proposed methodology, namely the physical constraints.

\subsection{Cross-validation and random seed}
In order to consider the variability due to the data used during optimization, each model configuration was trained on multiple cross-validation splits. When dealing with time series, it is important to design the cross-validation strategy in a way that ensures independence between sets. At the same time, in our setup it is desirable that samples of different sets are roughly equally distributed throughout the year. To do so, we opted for a simple four-folds cross-validation with a holdout set for testing. Specifically, of our five years of data we used 20\% for testing, 60\% for training and 20\% for validation, and we removed values between sets in a gap of five days in order to ensure independence. A total of four years was considered for training and validation, meaning each of the four cross-validation folds considered a different year in the dataset. The dataset partitioning is shown in the Supplementary Material's (SM) Fig. SM1. For each cross-validation split, we applied a standard normalization to the inputs of each set based on the training set's mean and standard deviation. Moreover, to account for the stochasticity of neural networks optimization (due to weights initialization and gradient descent), each model configuration was trained with 3 different random seeds. In total, for any given approach, 12 trials were conducted with the trained models and subsequently all evaluated on the same holdout test dataset.

\begin{table}
    \caption{List of predictors and predictands considered in this study.}
    \label{tab:mapping}
    \renewcommand{\arraystretch}{1.1} 
    \begin{tabular}{|l|l|l|}
        \topline
        Predictor & Units & Symbol\\
        \midline
        Air temperature ensemble average & °C & $T$\\
        Dew point temperature ensemble average & °C & $T_d$\\
        Dew point deficit ensemble average & °C & $T_{\mathrm{def}}$\\
        Relative humidity ensemble average & \% & $RH$ \\
        Surface air pressure ensemble average & hPa & $P$\\
        Water vapor mixing ratio ensemble average & g\,kg$^{-1}$ & $r$\\
        Forecast leadtime & h & \\
        Sine component of day of the year &  & \\
        Cosine component of day of the year &  & \\
        Sine component of hour of the day &  & \\
        Cosine component of hour of the day &  &\\
        \topline
        Predictand & Units & Symbol\\
        \midline
        2\,m Air temperature & °C & $T$\\
        2\,m Dew point temperature & °C & $T_d$\\
        2\,m Surface air pressure & hPa & $P$\\
        2\,m Relative humidity & \% & $RH$\\
        2\,m Water vapor mixing ratio & g\,kg$^{-1}$  & $r$\\
        \botline
    \end{tabular}
\end{table}

\subsection{Multi-task neural networks for postprocessing}

To keep our DL framework general, the basic building block used for all models in this study is a fully connected neural network (FCN, see Fig. \ref{fig:architectures}) that takes as inputs a vector containing the predictors in Table \ref{tab:mapping} and a vector with station IDs. Station IDs are mapped into a n-dimensional, real-valued vector $\textbf{z} \in \mathbb{R}^6$ via an embedding layer and concatenated with the predictors. This approach, here referred to as "unconstrained setting", is the same proposed by \citet{rasp_neural_2018} and may be regarded as state-of-the-art for the case of local postprocessing of surface variables \citep{schulz_machine_2022}. It is convenient because it allows for training a single model to do local postprocessing at many stations, instead of training one model for each station, making its operational implementation easier. We note that our $\mathbb{R}^6$ embedding has more dimensions than the $\mathbb{R}^2$ found in \citet{rasp_neural_2018}: this is likely due to the fact that we target five variables simultaneously and to the complex topography of the Alps (compared to German stations). The forecasts are deterministic, and we train a model to predict multiple target variables simultaneously. As such, we are dealing with a case of multi-task learning  \citep[see][for a survey on the subject]{crawshaw_multi-task_2020}, where the individual errors of each task contribute to the objective loss function. \citet{kendall_multi-task_2017} observed that the relative weighting of each task's loss had a strong influence on the overall performance of such models. This is fairly intuitive in our application because tasks have have different scales (due to different units) and uncertainties. The authors proposed a weighting scheme based on the homoscedastic uncertainty of each task, where the weights are learned during optimization. To the best of our knowledge, this approach for multi-task learning is new in the postprocessing field. If jointly postprocessing multiple meteorological parameters simultaneously becomes more common, it will be important to design optimal weighting schemes. We use the Mean Squared Error (MSE) for the loss function $\mathcal{L}_k$ of each task $k$. For a predicted value $\hat{y}_{k,i}$ in physical units (we will use the that notation for predicted values throughout the rest of the paper) and an observed value $y_{k,i}$, the task loss $\mathcal{L}_k$ is hence defined as:

\begin{equation}
    \mathcal{L}_k \overset{\mathrm{def}}{=}\sum_{i=1}^{N_{\mathrm{samples}}}\frac{\left(y_{k,i}-\hat{y}_{k,i}\right)^{2}}{N_{\mathrm{samples}}},
\end{equation}

where $N_{\mathrm{samples}} $ is the number of samples. For $p$ tasks, we define the combined loss $\mathcal{L} $ as: 

\begin{equation}
    \label{eq:wloss}
    \mathcal{L} \overset{\mathrm{def}}{=}  \sum_{k=1}^{p} \left[\frac{1}{2}\frac{\mathcal{L}_k}{\sigma_k^2}  +  \log{\sigma_k}\right].
\end{equation}

Each task's loss $\mathcal{L}_k$ is scaled by the homoscedastic uncertainty represented by $\sigma^2_k$, and a regularising term $\log{\sigma_k}$ is added to prevent degenerating towards a zero-weighted loss. In practice, for improved numerical stability, we learn the log variance $\eta_{k} \overset{\mathrm{def}}{=} \log{\sigma_{k}^2}$ for each task $k$ and Eq.\,\ref{eq:wloss} becomes:

\begin{equation}
    \label{eq:wloss2}
    \mathcal{L} =  \sum_{k=1}^{p}\frac{\mathcal{L}_k \exp\left(-\eta_{k}\right) + \eta_{k}}{2},
\end{equation}

where the division by 2 can be ignored for optimization purposes as it does not influence the minimization objective. To avoid having to learn large biases, we initialize the bias vector in the model's output layer using the training set-averaged output vector, which facilitates optimization. The optimization is insensitive to the initial values of this bias vector as long as it has the same order of magnitude as the mean output.

\subsection{Enforcing analytic constraints in neural networks}
The methodology used here follows \citet{beucler_enforcing_2021} and was first applied to neural networks emulating subgrid-scale parametrization for climate modeling. Conservation laws are enforced during optimization via constraints in the architecture or the loss function. In this study we aim to enforce dependencies between variables using the ideal gas law and an an approximate integral of the Clausius-Clapeyron equation used operationally at MeteoSwiss. Specifically, we postprocess air temperature ($T$, °C), dew point temperature ($T_d$, °C), surface air pressure ($P$, hPa), relative humidity ($RH$, \%) and water vapor mixing ratio ($r$, g\,kg$^{-1}$). We then aim to enforce the following constraints:

\begin{equation}
    \label{eq:con}
    \begin{aligned}
       RH = f(T, T_d) = \exp\left( \frac{a\cdot T_d}{b + T_d} - \frac{a\cdot T}{b + T} \right),\\
       r = g(P, T_d) = \frac{622.0 \cdot c\cdot \exp\left(\frac{a\cdot T_d}{b + T_d}\right)}{P - c\cdot \exp\left(\frac{a\cdot T_d}{b + T_d}\right)},
    \end{aligned}
\end{equation}

where $a$, $b$ and $c$ are empirical coefficients, as explained below. The system of interest includes five variables and two constraints, which leaves us with three degrees of freedom. The constraints functions $f$ and $g$ are derived from the following equations:

\begin{equation}
    \label{eq:tetens}
    e = c\cdot \exp\left(\frac{a\cdot T_d}{b + T_d}\right)\quad \text{   and   } \quad e_s = c\cdot \exp\left(\frac{a\cdot T}{b + T}\right),
\end{equation}

\begin{equation}
    \label{eq:rh}
    RH = \frac{e}{e_s} \cdot 100,
\end{equation}

\begin{equation}
    \label{eq:r}
    r = 1000 \cdot \frac{0.622 \cdot e}{p - e}.
\end{equation}

We note from Eq.\,\ref{eq:tetens} that the parameters of interest are linked by two additional physical quantities: the water vapor pressure ($e$, hPa) and the saturation water vapor pressure ($e_s$, hPa). Eq.\,\ref{eq:tetens} is structurally identical to the August–Roche–Magnus equation, an approximate integral of the Clausius-Clapeyron relation accurate for standard weather conditions. We use $a=17.368$, $ b=238.83$ and  $c=6.107$\,hPa for $T\geq0$;  $a=17.856$,  $b=245.52$, $c=6.108$\,hPa otherwise. We made this choice to ensure consistency with MeteoSwiss' internal processing of meteorological variables, but other values can be found in the literature \citep[e.g.][]{lawrence_relationship_2005}. Eq.\,\ref{eq:r} is a formula for water vapor mixing ratio derived from the ideal gas law for dry air and water vapor, and can be found in many common textbooks \citep[e.g.][]{emanuel_atmospheric_1994}. We multiply by 1000 to express $r $ in g\,kg$^{-1}$ rather than g\,g$^{-1}$.

\begin{figure*}[ht]
    \centerline{\includegraphics[width=\textwidth]{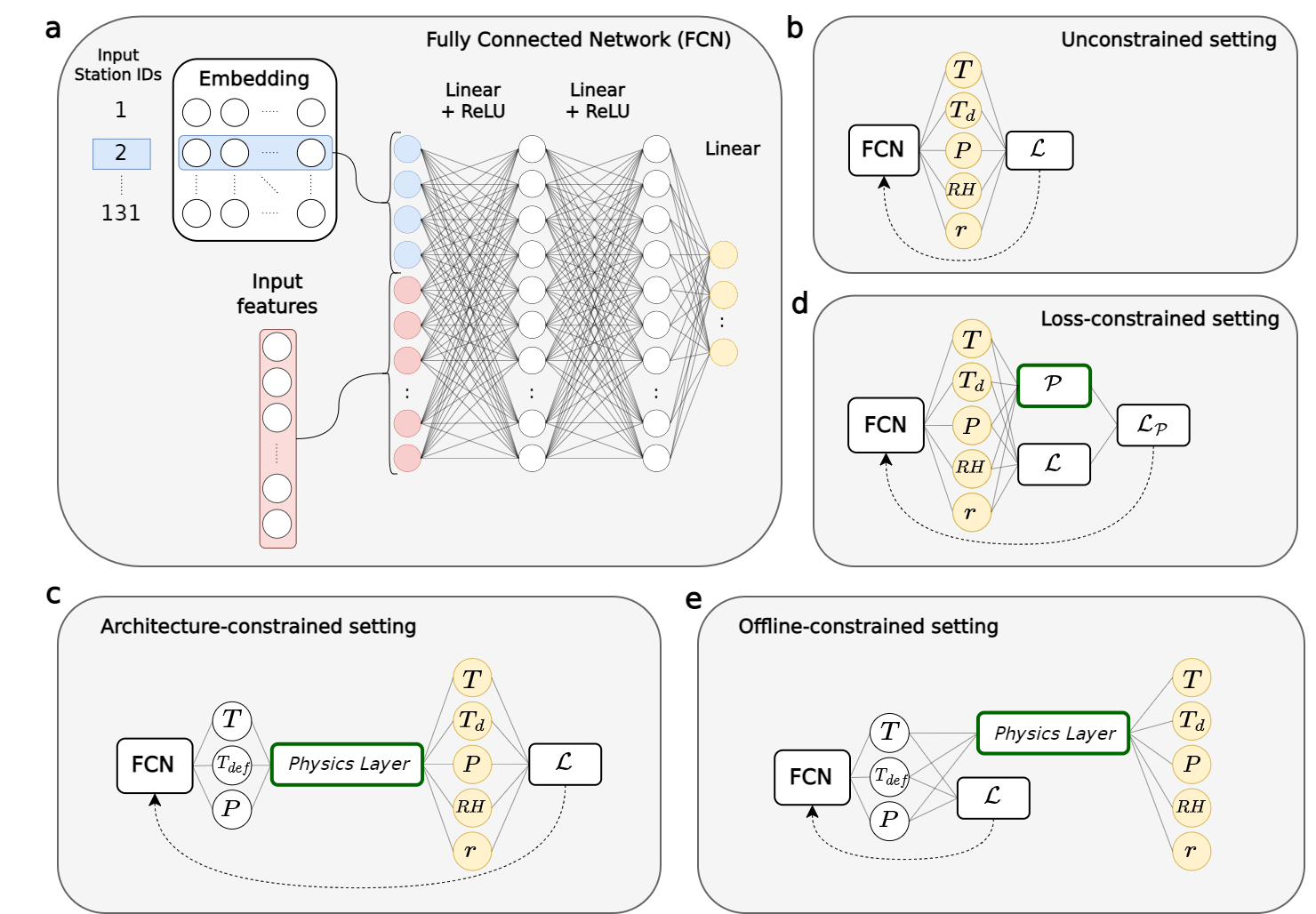}}
    \caption{Summary of the models used in this study. \textbf{a}, the basic building block of all models, a fully connected network (FCN) preceded by an embedding layer. \textbf{b}, the unconstrained setting used as a baseline, where all target variables are predicted directly. \textbf{c}, the architecture constrained setting, including a physical constraints layer that takes a subset of the target variables as inputs and returns the complete prediction. \textbf{d}, the loss constrained neural network, in which physical consistency is enforced by adding a physics-based penalty $\mathcal{P}$ to the conventional loss $\mathcal{L}$. \textbf{e}, an offline-constrained neural network, where constraints are only applied \textit{after} training using the constraints layer.}
    \label{fig:architectures}
\end{figure*}

We proceed to implement and compare two approaches how to enforce physical constrains in our networks:
\paragraph{architecture-constrained setting:}
the constraints are enforced by using a layer that uses equations~\ref{eq:con} to derive $RH$ and $r$ from $T$, $T_{\mathrm{def}}$ and $P$. $T_{\mathrm{def}}$ is the dew point deficit, to which we apply a Rectified Linear Unit (ReLU) activation function to ensure positivity before computing dew point temperature as $T_d\overset{\mathrm{def}}{=}T-T_{\mathrm{def}}$. This additional step allows us to enforce that $T \geq T_d$ and $RH \in [0,100]$, two desirable properties in this case. We show the constrained architecture in Figure \ref{fig:architectures}c. The trainable part of the model has the same number of layers and units as the unconstrained architecture, but directly predicts only a subset of variables. In our case, which has 5 outputs and 2 constraints, we directly predict 3 variables and derive the last 2 via a custom-defined layer encoding equations~\ref{eq:con}. An important point is that the choice of which variables are computed first and which are derived analytically is arbitrary. Given $n=5$ variables and $q=2$ constraints, the total number of possibilities in our case is $\frac{n!}{q!(n-q)!} = 10$. Nevertheless, there are differences in the actual implementation that point in favor of some configurations. For example, one may want the analytic constraints arranged in a way such that numerical stability is not a concern, e.g., avoiding division by zero or asymptotes of logarithmic functions.

\paragraph{loss-constrained setting:}
the constraints are enforced by using an additional physics-based loss term that includes a penalty term $\mathcal{P}$ based on residuals from our set of analytic equations. As for the architecture-constrained approach, the variables that we choose to calculate the residuals are arbitrary. Based on Eq.~\ref{eq:con}, we define the following constraints:

\begin{equation}
    \begin{aligned}
        \left\{
        \begin{array}{*{3}{r}l}
          \mathcal{P}_{RH} \overset{\mathrm{def}}{=} \hat{RH} - f(\hat{T}, \hat{T_d}) = 0,\\
           \mathcal{P}_r \overset{\mathrm{def}}{=} \hat{r} - g(\hat{P}, \hat{T_d}) = 0,
        \end{array}
        \right.
    \end{aligned}
\end{equation}

where physical violations result in non-zero residuals. Using the L2-norm for consistency with our MSE loss, we formulate the penalty term $\mathcal{P}$ used in the loss function as:

\begin{equation}
    \mathcal{P} \overset{\mathrm{def}}{=} \frac{(\mathcal{P}_{RH})^2}{\sigma^2_{RH}} +  \frac{(\mathcal{P}_r)^2}{\sigma^2_r},
\end{equation}

where we square the residuals $\mathcal{P}_{RH}$ and $\mathcal{P}_r$ to penalize larger violations more, and then scale by the variance of the observed values $\sigma^2_{RH}$ and $\sigma^2_r$ in order to normalize the contribution of the two terms. Finally, the physical penalty term is added to the conventional loss and our training objective becomes minimizing the physically-constrained loss function $\mathcal{L}_{\mathcal{P}} $:

\begin{equation}
    \mathcal{L}_{\mathcal{P}} \overset{\mathrm{def}}{=}  (1-\alpha) \mathcal{L} + \alpha \mathcal{P},
\end{equation}

where $\alpha \in \left[0,1\right]$ is a hyperparameter used to scale the contribution of the physical penalty term. Note that in contrast with the \textit{hard} constraints in the architecture, with the \textit{soft} constraints in the loss, we have no guarantee that $\mathcal{P}=0$ because stochastic gradient descent does not generally lead to a zero loss. 

Finally, to assess whether enforcing constraints \textit{during} optimization is advantageous, we additionally introduce an ``offline-constrained'' setting (see Figure \ref{fig:architectures}e) in which constraints are enforced \textit{after} training. Specifically, we train a model to minimize the MSE of $T$, $T_d$ (derived from $T$ and $T_{\mathrm{def}}$) and $P$. $RH$ and $r$ are then calculated \textit{after} training so as to exactly enforce our physical constraints.

\subsection{Libraries, hyperparameters and training}
We use the PyTorch deep learning library \citep{paszke_pytorch_2019} to implement our models, the Ray Tune library \citep{liaw_tune_2018} for hyperparameter tuning, and Snakemake \citep{molder_sustainable_2021} to manage our workflow \footnote{code at \url{www.github.com/frazane/pcpp-workflow}}. For training we use the Adam optimizer \citep{kingma_adam_2014} with the exponential decay rates set to $\beta_1=0.99$ and $\beta_2=0.999$ for the first and second moments \citep{kingma_adam_2014}, implement an early-stopping rule based on a validation loss to avoid overfitting. We use the aggregated normalized mean absolute error (NMAE) aggregated over all 5 outputs as our validation loss:

\begin{equation}
    \mathrm{NMAE} \overset{\mathrm{def}}{=} \frac{1}{5}\sum_{k=1}^{5}\sum_{i=1}^{N_{\mathrm{samples}}}\frac{|y_{i,k} - \hat{y_{i,k}}|}{\sigma_{k}},
\end{equation}

and halt training after 5 epochs in the absence of improvements in the validation loss. This metric was chosen for the validation and early stopping because it proved to be more robust to sudden model changes during training, compared to the training loss. We save the model state with the lowest validation loss of all training epochs. The hyperparameters used to produce the main results of this study are shown in Table \ref{tab:hp}. We chose them after running a hyperparameter tuning algorithm for the unconstrained model that considered the aggregated loss of all cross-validation splits. The best performing hyperparameters configuration was then applied to all models. The loss-constrained model also required a hyperparameter to scale the influence of the physics-based penalty term. After testing different values, $\alpha $ was set to 0.995. We discuss this choice in the next section.

\begin{table}
    \caption{Hyperparameters used to train the models, along with their optimal value and the search space used for tuning the unconstrained model, represented as the range or set of possible values followed by the sampling method. After selecting the five best configurations automatically, we chose the one with the lowest number of trainable parameters for parsimony.}
    \renewcommand{\arraystretch}{1.1} 
    \begin{tabular*}{\hsize}{@{\extracolsep\fill}lll@{}}
        \topline
        \textbf{Parameter}      &       \textbf{Value}  &   \textbf{Search} \\
        \midline
        Learning rate           &       0.0007           &   [0.01-0.001] $\sim{}$ loguniform  \\
        Batch size              &       512            &   \{256, 512, 1024, 2048\} $\sim{}$  choice \\
        Units in layer 1                &       256             &   \{32, 64, 128, 256\} $\sim{}$  choice     \\
        Units in layer 2                &       256              &   \{32, 64, 128, 256\} $\sim{}$  choice     \\
        Embedding               &       6               &   \{2, 3, 4, 5, 6\} $\sim{}$  choice     \\
        Patience                &       5 epochs        &                                       \\
        $\alpha$ (loss-constrained) &   0.995      &                                       \\
        \botline
    \end{tabular*}
    \label{tab:hp}
\end{table}

\section{Results and discussion\label{sec:results}}

In this section, we present the results of our models when evaluated on unseen data. We will first compare the performance and physical consistency of different architectures (Sec.~\ref{sec:results}\ref{sub:main_results}) before discussing data efficiency (Sec.~\ref{sec:results}\ref{sub:data_efficiency}) and generalization ability (Sec.~\ref{sec:results}\ref{sub:generalization}).

\subsection{Predictive performance and physical consistency\label{sub:main_results}}

\begin{figure*}[ht]
    \centerline{\includegraphics[width=\textwidth]{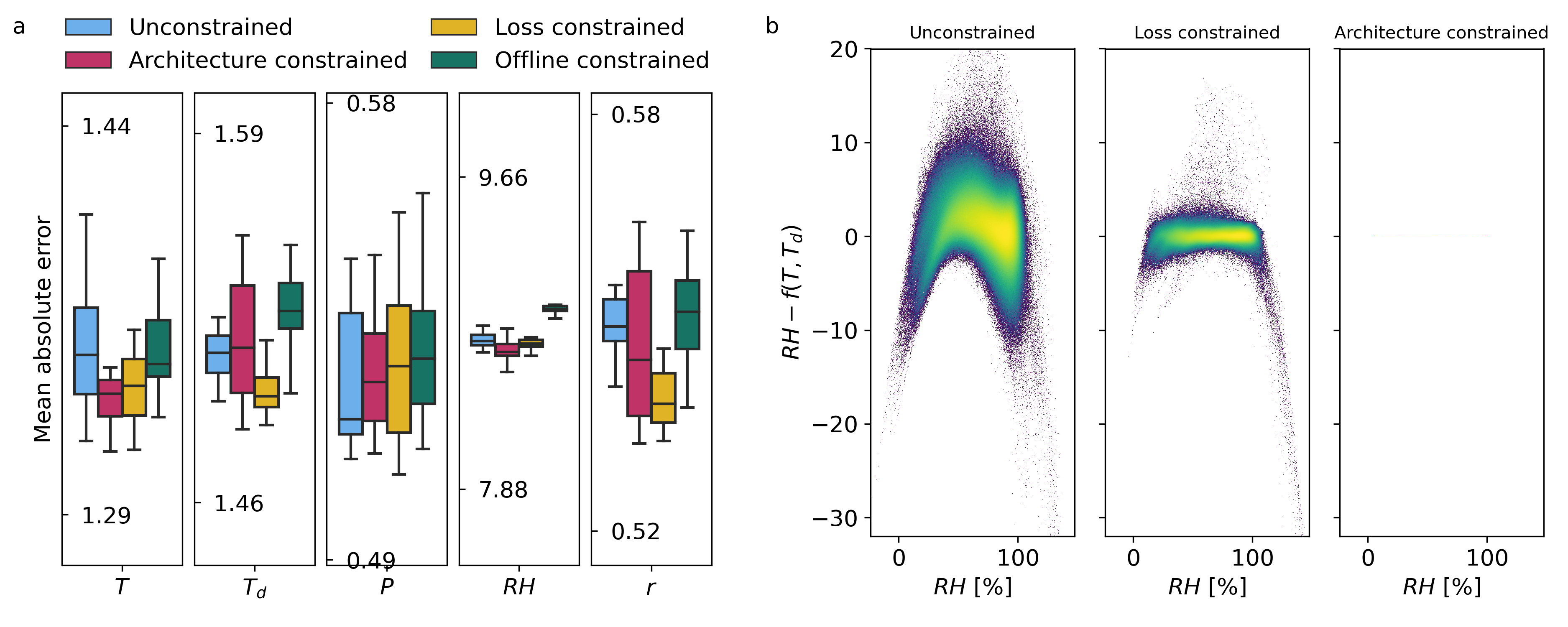}}
    \caption{\textbf{a}, MAE for each target variable and approach, where the boxplot distribution represents the 9 trials using different cross-validation splits and random seeds. Note that the ranges of these distributions are relatively small compared to the absolute values of the error metric. \textbf{b}, Scatter plot representing the distribution of physical violations $\mathcal{P}_{RH}$ in RH units, as a function of $RH$, using all samples of all trials, where points are color-coded by density. Physical violations are deviations from the zero line.}
    \label{fig:mae_boxplots}
\end{figure*}

We use two metrics to evaluate the overall performance of our models: the mean absolute error (MAE) and the Mean Squared Skill Score (MSSS) calculated with respect to the raw NWP forecast, defined as:

\begin{equation}
    \mathrm{MSSS} = 1 - \frac{\mathrm{MSE_{PP}}}{\mathrm{MSE_{NWP}}},
\end{equation}

where $\mathrm{MSE_{PP}}$ and $\mathrm{MSE_{NWP}}$ represent the MSE for our postprocessed forecasts and the NWP forecast respectively. The MSSS presented here was first computed on each station individually and then averaged.
The overall results are presented in Table \ref{tab:res1} and Figure \ref{fig:mae_boxplots}a for each variable. For both MAE and MSSS these results show that the performance is comparable for all NN architectures. There are small differences in performance of each setting on the different tasks, which we hypothesize are related to the influence of the constraints coupled with the multi-task loss weighting. Overall, we observe slightly better results for the loss and architecture constrained approaches. Table \ref{tab:count1} shows the results of the Diebold-Mariano predictive performance tests for the MAE (we followed the implementation of \cite{schulz_machine_2022}) applied to each station and leadtime individually: the reported values are the percentage of tests for which a method significantly outperformed another method. Overall, these results appear to reflect those from Table \ref{tab:res1}: we don't observe a clear winner, but the offline constrained approach appears to be generally worse than the others with the exception of pressure. While we did not focus on obtaining the best possible performance for our models, it is useful to know how other widely known postprocessing approaches perform as reference: in the case of air temperature, we obtained a MAE of 1.63\,°C for an altitude-corrected NWP forecast (using a fixed lapse rate of 6\,°C km$^{-1}$), and found a value of 1.01\,°C for the MOSMIX product provided by Deutscher Wetterdienst\footnote{https://www.dwd.de/}, noting that this product also includes real-time information from ground stations. In comparison, our models exhibit MAEs of approximately 1.35\,°C.
We note that the MSSS values are surprisingly high, especially for pressure. These high MSSS values are due to the large errors in the NWP model. For variables that are strongly tied to elevation, such as pressure and temperature, the differences in the NWP model elevation and the true station elevation result in consistently large biases. These elevation differences can be larger than 100 meters. Taking the example of pressure, the mean bias at certain stations is almost 90\,hPa, which is reduced to almost 0\,hPa by the postprocessing models, explaining the high MSSS values.

\begin{table*}
    \caption{Two performance metrics: The mean-absolute error (MAE, lower values are better) and mean squared skill score (MSSS, higher values are better) for each considered NN architecture (rows) and target variables (columns), averaged over the test set.}
    \renewcommand{\arraystretch}{1.1} 
    \begin{tabular*}{\hsize}{@{\extracolsep\fill}l lllll lllll@{}}
        \topline
        &\multicolumn{2}{c}{$T$} & \multicolumn{2}{c}{$T_d$} & \multicolumn{2}{c}{$P$} & \multicolumn{2}{c}{$RH$} & \multicolumn{2}{c}{$r$}\\ 
        \cmidrule(lr){2-3} \cmidrule(lr){4-5} \cmidrule(lr){6-7} \cmidrule(lr){8-9} \cmidrule(lr){10-11} 
        & MAE & MSSS & MAE & MSSS & MAE & MSSS & MAE & MSSS & MAE & MSSS \\ 
        Unconstrained            & 1.355          & 0.351          & 1.514          & 0.318          & 0.529          & 0.618          & 8.737          & 0.290          & 0.550          & 0.252          \\ 
        Architecture constrained & \textbf{1.340} & \textbf{0.357} & 1.517          & 0.305          & \textbf{0.527} & \textbf{0.624} & \textbf{8.677} & 0.286          & 0.547          & 0.247          \\ 
        Loss constrained         & 1.343          & 0.356          & \textbf{1.501} & \textbf{0.326} & 0.530          & 0.623          & 8.722          & \textbf{0.297} & \textbf{0.539} & \textbf{0.274} \\ 
        Offline constrained      & 1.356          & 0.341          & 1.530          & 0.295          & 0.533          & 0.607          & 8.911          & 0.261          & 0.551          & 0.238          \\ 
        \\ 
   \end{tabular*}
   \label{tab:res1}
\end{table*}

\begin{table*}
    \footnotesize
    \caption{For each target variable, the percentage of tests for which an approach (on a row) produces significantly better forecasts than another (on a column), according to Diebold-Mariano statistical tests performed with the MAE. Reported is also on the average percentage of "wins" or "losses" for each approach. Tests are applied to each station and leadtime individually.}
    \renewcommand{\arraystretch}{1.1}
    \begin{subtable}{1\textwidth}
    \caption{Air temperature}
    \begin{tabular*}{\hsize}{@{\extracolsep\fill}l llll l@{}}
        \topline
        & Unconstrained & Architecture constrained & Loss constrained & Offline constrained & Winning average\\ 
        \midline
        Unconstrained &  & 6.72 & 4.91 & 15.67 & 9.10 \\ 
        Architecture constrained & 32.42 &  & 16.61 & 34.13 & 27.72 \\ 
        Loss constrained & 26.75 & 12.69 &  & 33.33 & 24.26 \\ 
        Offline constrained & 14.32 & 6.83 & 8.98 &  & 10.04 \\ 
        \midline
        Losing average & 24.50 & 8.75 & 10.17 & 27.71 &  \\ 
        \\ 
   \end{tabular*}
   \end{subtable}
    \begin{subtable}{1\textwidth}
    \caption{Dew point temperature}
    \begin{tabular*}{\hsize}{@{\extracolsep\fill}l llll l@{}}
        \topline
        & Unconstrained & Architecture constrained & Loss constrained & Offline constrained & Winning average\\ 
        \midline
        Unconstrained &  & 12.87 & 6.47 & 22.86 & 14.07 \\ 
        Architecture constrained & 6.18 &  & 3.82 & 17.92 & 9.31 \\ 
        Loss constrained & 26.32 & 33.62 &  & 49.07 & 36.34 \\ 
        Offline constrained & 2.29 & 1.85 & 1.27 &  & 1.81 \\ 
        \midline
        Losing average & 11.60 & 16.12 & 3.85 & 29.95 &  \\ 
        \\ 
   \end{tabular*}
   \end{subtable}
    \begin{subtable}{1\textwidth}
    \caption{Surface air pressure}
    \begin{tabular*}{\hsize}{@{\extracolsep\fill}l llll l@{}}
        \topline
        & Unconstrained & Architecture constrained & Loss constrained & Offline constrained & Winning average\\ 
        \midline
        Unconstrained &  & 28.50 & 30.61 & 32.13 & 30.41 \\ 
        Architecture constrained & 32.82 &  & 36.28 & 38.75 & 35.95 \\ 
        Loss constrained & 32.82 & 28.10 &  & 36.17 & 32.36 \\ 
        Offline constrained & 22.90 & 22.43 & 28.28 &  & 24.54 \\ 
        \midline
        Losing average & 29.52 & 26.34 & 31.72 & 35.68 &  \\ 
        \\ 
   \end{tabular*}
   \end{subtable}
    \begin{subtable}{1\textwidth}
    \caption{Relative humidity}
    \begin{tabular*}{\hsize}{@{\extracolsep\fill}l llll l@{}}
        \topline
        & Unconstrained & Architecture constrained & Loss constrained & Offline constrained & Winning average\\ 
        \midline
        Unconstrained &  & 9.63 & 11.89 & 38.10 & 19.87 \\ 
        Architecture constrained & 21.12 &  & 19.81 & 57.00 & 32.64 \\ 
        Loss constrained & 17.59 & 14.07 &  & 45.91 & 25.86 \\ 
        Offline constrained & 8.65 & 1.82 & 5.74 &  & 5.40 \\ 
        \midline
        Losing average & 15.79 & 8.51 & 12.48 & 47.00 &  \\ 
        \\ 
   \end{tabular*}
   \end{subtable}
    \begin{subtable}{1\textwidth}
    \caption{Water vapor mixing ratio}
    \begin{tabular*}{\hsize}{@{\extracolsep\fill}l llll l@{}}
        \topline
        & Unconstrained & Architecture constrained & Loss constrained & Offline constrained & Winning average\\ 
        \midline
        Unconstrained &  & 7.74 & 1.45 & 10.58 & 6.59 \\ 
        Architecture constrained & 20.39 &  & 4.98 & 19.12 & 14.83 \\ 
        Loss constrained & 57.76 & 32.82 &  & 45.69 & 45.43 \\ 
        Offline constrained & 6.47 & 3.96 & 1.09 &  & 3.84 \\ 
        \midline
        Losing average & 28.21 & 14.84 & 2.51 & 25.13 &  \\ 
        \\ 
   \end{tabular*}
   \end{subtable}
   \label{tab:count1}
\end{table*}

Figure \ref{fig:mae_boxplots}b depicts the physical consistency of the predictions. On the vertical axis is $\mathcal{P}_{RH}$, which is the difference between the predicted $RH$ and its physically-consistent counterpart derived from the constraint function $f(T, T_d)$ (Eq.~\ref{eq:con}), while we show predicted values on the horizontal axis. Deviations from zero are therefore considered physical violations, as well as values that are not between 0 and 100. Compared to the unconstrained approach, we observe a noticeable decrease of violations using the loss-constrained approach, although large violations still occur at the ends of the RH distribution, where values larger than 100\% still occur. The architecture-constrained models bring physical violations to zero to within machine precision. These results are consistent with \citet{beucler_enforcing_2021}. As a side remark, we note that in the unconstrained approach larger violations tend to occur more at the tails of the distribution, which could indicate that it is more difficult to converge to a physically consistent solution if the samples are scarce.

\begin{figure}[ht]
    \centerline{\includegraphics[width=19pc]{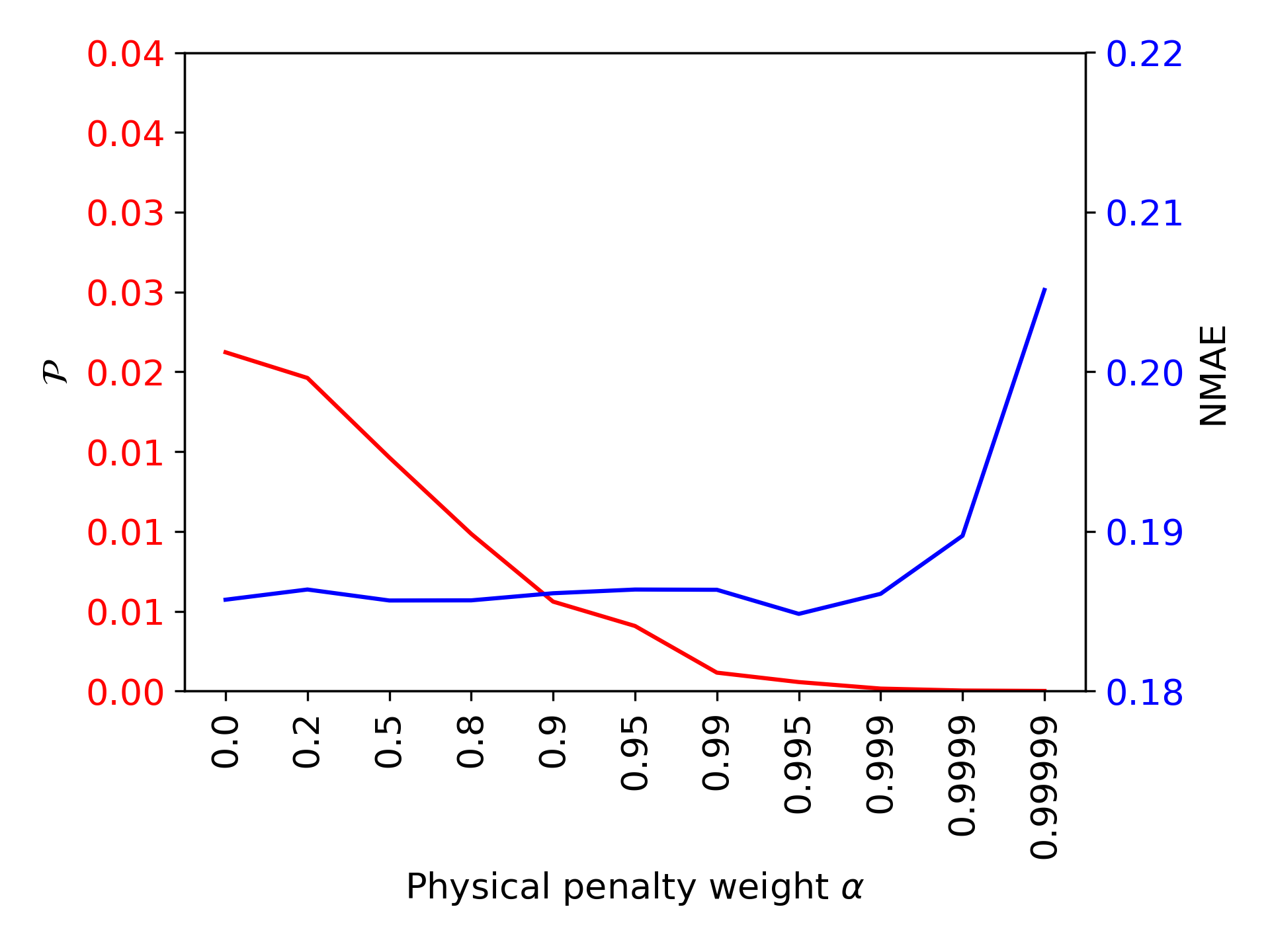}}
    \caption{Effect of the hyperparameter $\alpha$ on both the overall physical violation $\mathcal{P}$ (in red) and the NMAE (in blue) for the test dataset.}
    \label{fig:alpha_choice}
\end{figure}

In order to choose an optimal value for the $\alpha$ hyperparameter, we tested several values and compared both the NMAE and the physical consistency of the predictions. The results are shown in Figure \ref{fig:alpha_choice} for the following values: $\alpha\in\{0., 0.2, 0.5, 0.8, 0.9, 0.95, 0.99, 0.995, 0.999, 0.9999, 0.99999\}$. We observe that the trade-off between physical constraints and performance is non-linear: up to $\alpha = 0.995$, there is little to no drawback in terms of performance. In contrast, for higher values of $\alpha $, the NMAE starts to increase. We note two things: first, the choice of this hyperparameter $\alpha $ can be chosen based on how much one wishes to prioritize physical consistency over error reduction. Second, one should be aware that the choice of the learning rate has a significant influence on $\alpha$'s impact (and vice-versa) and thus on this trade-off, although this was not further investigated in this study. We limit ourselves to observe that from a practical standpoint, this relationship is inconvenient as it makes model selection harder, and we consider it to be a drawback of the loss-constrained approach. 

\subsection{Robustness to data scarcity\label{sub:data_efficiency}}

\begin{figure*}[ht]
    \centerline{\includegraphics[width=\textwidth]{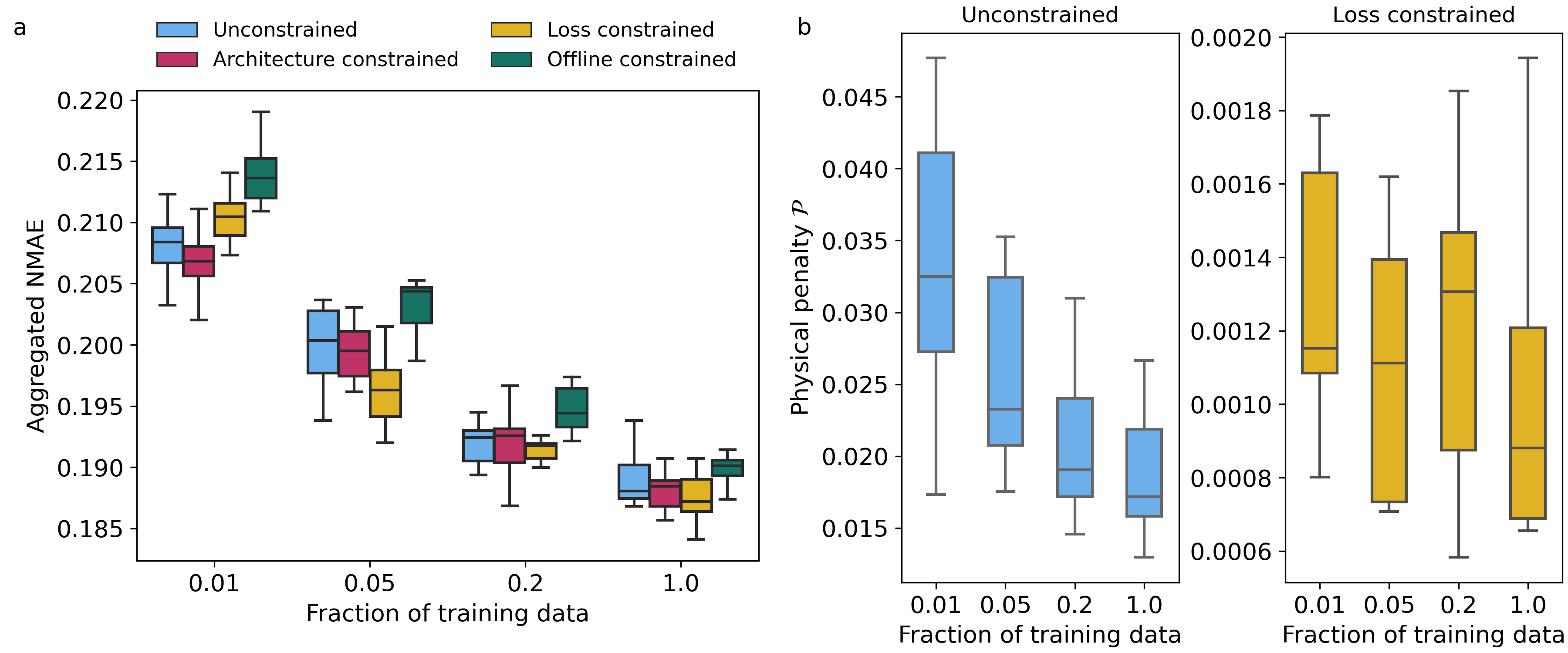}}
    \caption{\textbf{a}, NMAE for each reduction and approach, where the boxplot distribution represents the 9 trials using different cross-validation splits and random seeds. As the size of the training dataset reduces, constrained models perform relatively better. \textbf{b}, Box plot showing the physical penalty term $\mathcal{P}$'s distribution as a function of training data size for the unconstrained and loss constrained settings. The architecture- and offline-constrained approaches have zero penalty by construction.}
    \label{fig:data_efficiency}
\end{figure*}

Among the potential advantages of constraining neural networks using physical knowledge are improved robustness to data scarcity. The rationale is that because we reduce the hypothesis space of the model to a subset of physically-consistent solutions, we could expect the physics-constrained models to learn with fewer training samples, or to require fewer parameters. We have therefore designed an experiment in which we re-trained all models (with fewer parameters in order to reduce the chance of overfitting, see Table SM2) on increasingly reduced training datasets, namely 20\%, 5\% and 1\% of the full dataset. The reduction is applied by station, in order to ensure that all stations are still equally represented in the dataset. For instance, with the 1\% reduction we trained with roughly 200 samples for each station. The results, shown in Figure \ref{fig:data_efficiency}a, seem to indicate a relatively smaller decrease in performance for the architecture-constrained approach when the data is scarce, although this difference is rather small. Importantly, the added value of enforcing physical consistency in data scarce situations is emphasized by Figure \ref{fig:data_efficiency}b. For the unconstrained model in particular, the physical inconsistencies increase as we reduce the number of training samples. Conversely, for the architecture constrained approach, physical inconsistencies are always zero by construction.

\subsection{Generalization ability\label{sub:generalization}}
\begin{figure}[ht]
    \centerline{\includegraphics[width=19pc]{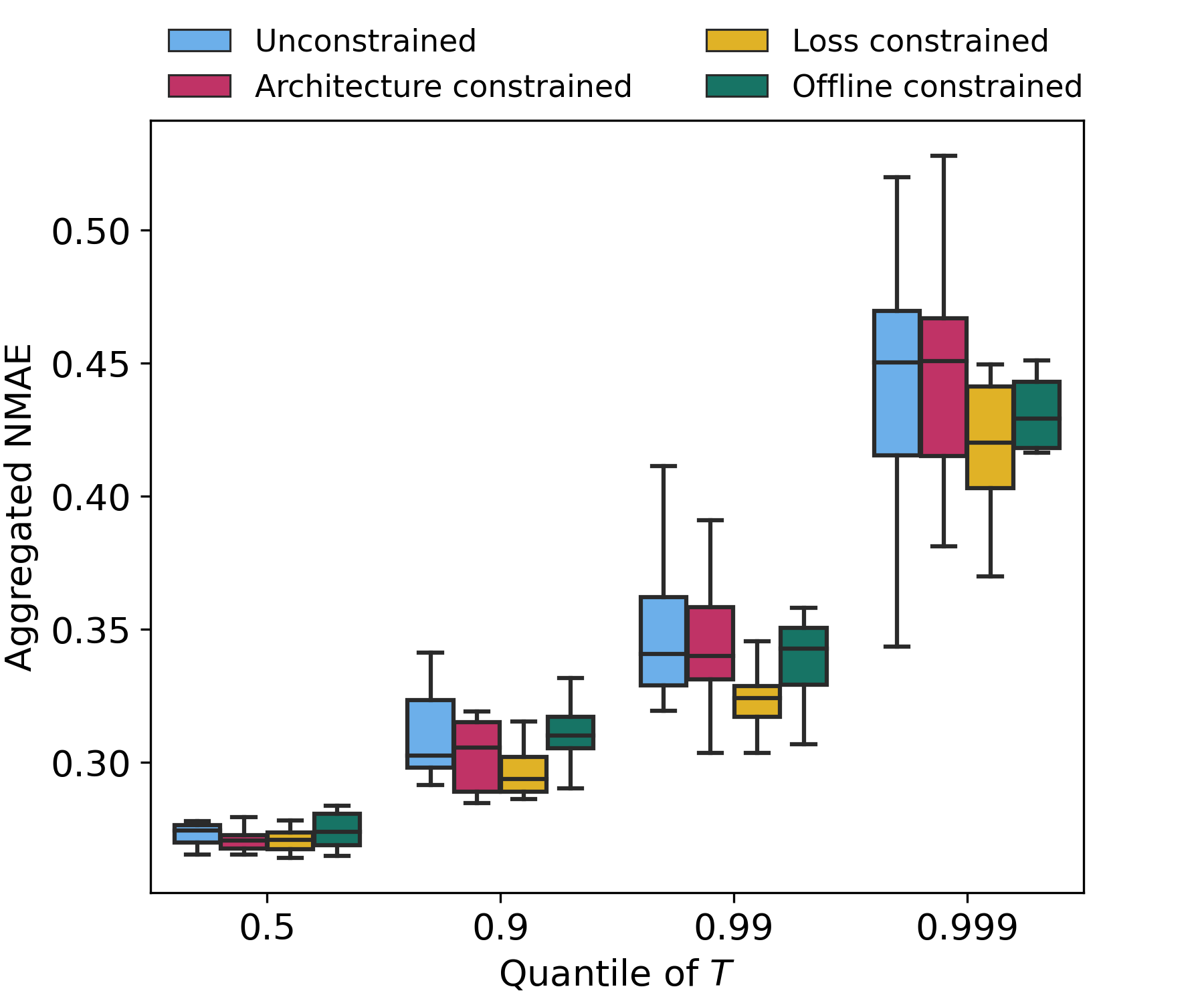}}
    \caption{NMAE for the test dataset containing samples from JJA, conditioned on different quantiles of the univariate temperature distribution. As expected, the error increases as temperature are more extremes, but the relative performance of the considered architectures does not change significantly.}
    \label{fig:generalization}
\end{figure}

A common finding of physics-informed ML is that physically-constraining models could help them generalize to unseeen conditions \citep{willard_integrating_2022}. In order to test the ability of our models to generalize to unseen weather situations, we design an experiment in which models are trained on a dataset that excludes the warm season (JJA), and then tested on the warm season only. This choice was motivated by the increasing relevance of record-shattering heat extremes in a warming climate. In such situations, the robustness of postprocessing models is put to test as they have to process and predict values never seen during training. We present our results in Figure \ref{fig:generalization}, and observe that physical constraints do not seem to impact the generalization capabilities of the model to unseen temperature extremes. This result, consistent with \cite{beucler_towards_2020}, suggests that the constraints of Equation \ref{eq:con} are insufficient to guarantee generalization capability for our mapping.

\section{Conclusion and outlook}

In this study, we have adapted a physically-constrained, deep learning framework to postprocess weather forecasts, which is new to our knowledge. More generally, we demonstrated simple ways to integrate scientific expertise, in the form of analytic equations, into a DL-based postprocessing model. Compared to unconstrained or loss-constrained models, architecture-constrained models enforce physical consistency to within machine precision without degrading the performance of the considered variables. We have also shown that physical constraints yield better predictions when data are scarce because of the increased value of physical consistency. However, we did not observe a significant advantage in terms of generalization capabilities. To interpret these results, it is useful to distinguish the data efficiency and the generalization experiment by their underlying challenge, that is, interpolation and extrapolation, respectively. Physically constraining outputs can help the model better interpolate data, but cannot mitigate the well-known limitations of neural networks when it comes to out-of-distribution inputs (extrapolation).

We believe that a significant value of the proposed methodology lies in its simplicity: any kind of equation, as long as it is differentiable, can be included in DL-based postprocessing models. Importantly, this extends beyond the context of meteorology and physics-based constraints, as we could easily imagine a similar methodology used to satisfy a diverse set of constraints defined by the end users. For future research on this topic we foresee (i) an extension to probabilistic forecasting, e.g. by adopting a generative approach for the creation of physically consistent ensembles; and (ii) an extension to a global postprocessing setup, where the model generalizes in space. Finally, an open question is whether physical constraints have a stronger effect in more challenging tasks, e.g. with higher-dimensional mappings or more marked non-linearities.  

%

%

\acknowledgments
We thank the members of APPP team at MeteoSwiss for helpful comments and feedback that significantly helped the project. FZ is supported by MeteoSwiss and ETH Zürich, DN and ML are supported by MeteoSwiss, while TB is supported by the Canton of Vaud in Switzerland. We also thank the Swiss National Supercomputing Centre (CSCS) for computing infrastructure.

%
%
\datastatement
The project's GitHub repository is accessible at \url{https://github.com/frazane/pcpp-workflow}. The raw data used to train the models is free for research and education purposes, and can be accessed via the IDAweb portal at \url{https://www.meteoswiss.admin.ch/services-and-publications/service/weather-and-climate-products/data-portal-for-teaching-and-research.html}. 

%





%



\bibliographystyle{ametsocV6}
\bibliography{references}

\end{document}


\maketitle

\begin{figure*}[ht]
    \centerline{\includegraphics[width=\textwidth]{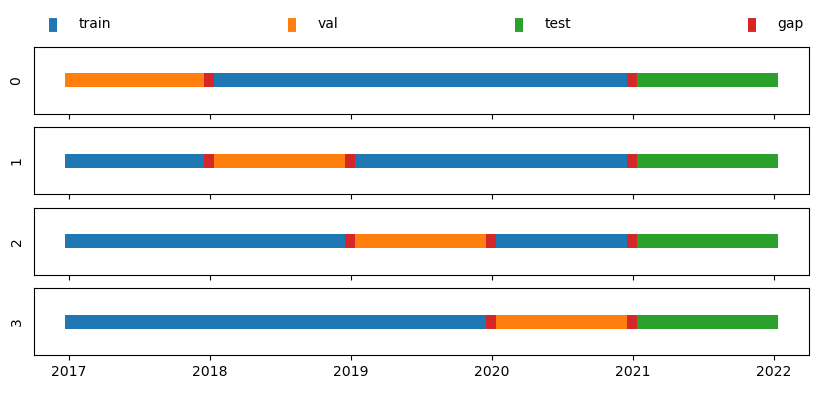}}
    \caption{Representation of the cross-validation partitioning strategy used in this study, with each row representing a single cross-validation split. The gap length is 5 days, although the rendering makes it appear larger.}
    \label{annex:cvsplit}
\end{figure*}

\begin{table*}
    \caption{Performance comparison table for the training set.}
    \renewcommand{\arraystretch}{1.1} 
    \begin{tabular*}{\hsize}{@{\extracolsep\fill}l lllll lllll@{}}
        \topline
        &\multicolumn{2}{c}{$T$} & \multicolumn{2}{c}{$T_d$} & \multicolumn{2}{c}{$P$} & \multicolumn{2}{c}{$RH$} & \multicolumn{2}{c}{$r$}\\ 
        \cmidrule(lr){2-3} \cmidrule(lr){4-5} \cmidrule(lr){6-7} \cmidrule(lr){8-9} \cmidrule(lr){10-11} 
        & MAE & MSSS & MAE & MSSS & MAE & MSSS & MAE & MSSS & MAE & MSSS \\ 
        Unconstrained & 1.304 & 0.513 & 1.433 & 0.502 & 0.499 & 0.963 & 8.442 & 0.451 & 0.537 & 0.474 \\ 
        Architecture constrained & 1.300 & 0.514 & 1.427 & 0.504 & 0.489 & 0.963 & 8.326 & 0.456 & 0.533 & 0.479 \\ 
        Loss constrained & 1.305 & 0.511 & 1.432 & 0.499 & 0.492 & 0.963 & 8.488 & 0.447 & 0.534 & 0.477 \\ 
        Offline constrained & 1.305 & 0.511 & 1.452 & 0.488 & 0.492 & 0.963 & 8.617 & 0.426 & 0.544 & 0.458 \\ 
        \\ 
   \end{tabular*}
\end{table*}

\begin{table}[]
    \caption{Hyperparameters chosen for the data scarcity experiment.}
    \renewcommand{\arraystretch}{1.1} 
    \begin{tabular*}{\hsize}{l @{\extracolsep\fill}lll@{}}
        \topline
        &\multicolumn{3}{c}{\textbf{Reduction}}\\
        \cmidrule(lr){2-4}
        \textbf{Parameters} &    0.2 & 0.05 & 0.01 \\
        \midrule
        Learning rate    & 0.004 & 0.0035 & 0.003  \\     
        Batch size       & 64  & 64  & 64   \\     
        Units in layer 1 & 64   & 64   & 32    \\     
        Units in layer 2 & 64   & 32   & 32    \\     
        Embedding        & 3    & 3    & 3     \\     
    \end{tabular*}
    
    \label{tab:ds_hp}
\end{table}

\begin{figure*}[ht]
    \centerline{\includegraphics[width=\textwidth]{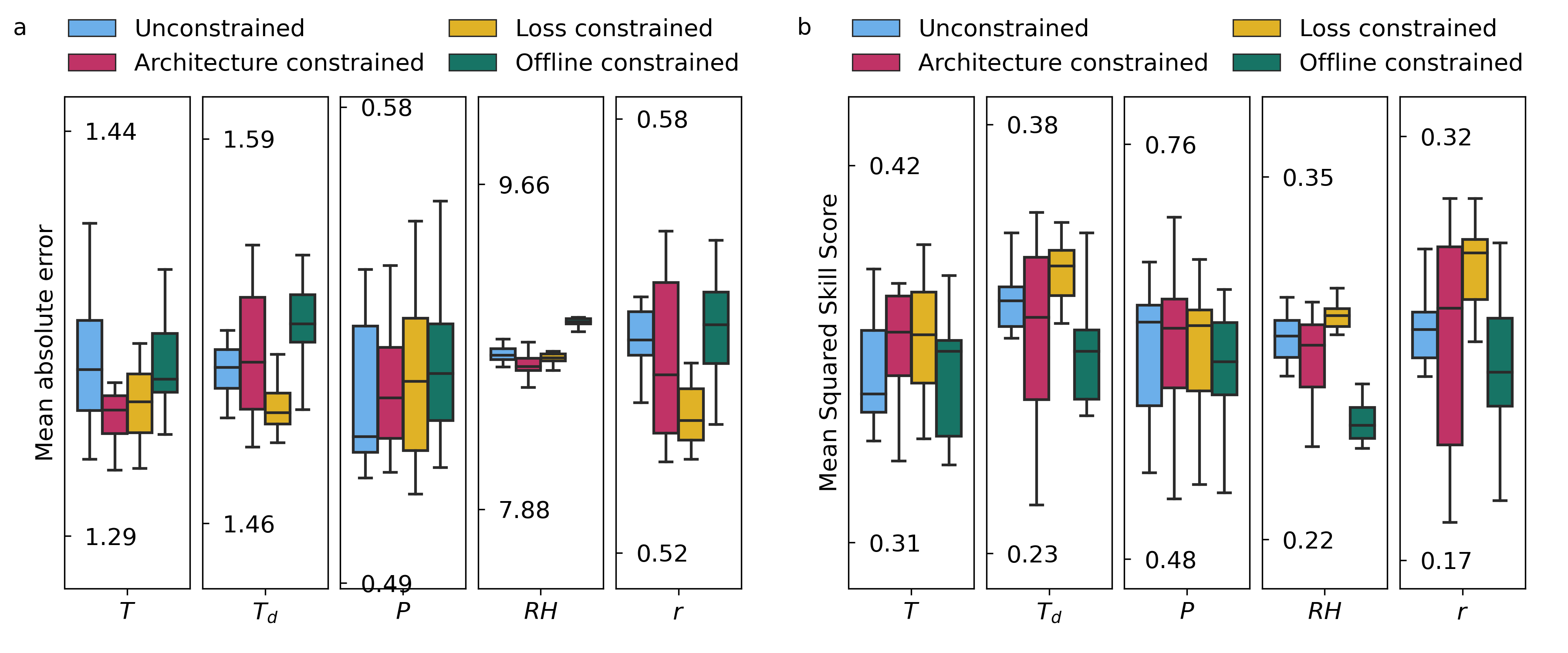}}
    \caption{Boxplots for the MAE and MSSS on the test set}
\end{figure*}

\begin{figure*}[ht]
    \centerline{\includegraphics[width=\textwidth]{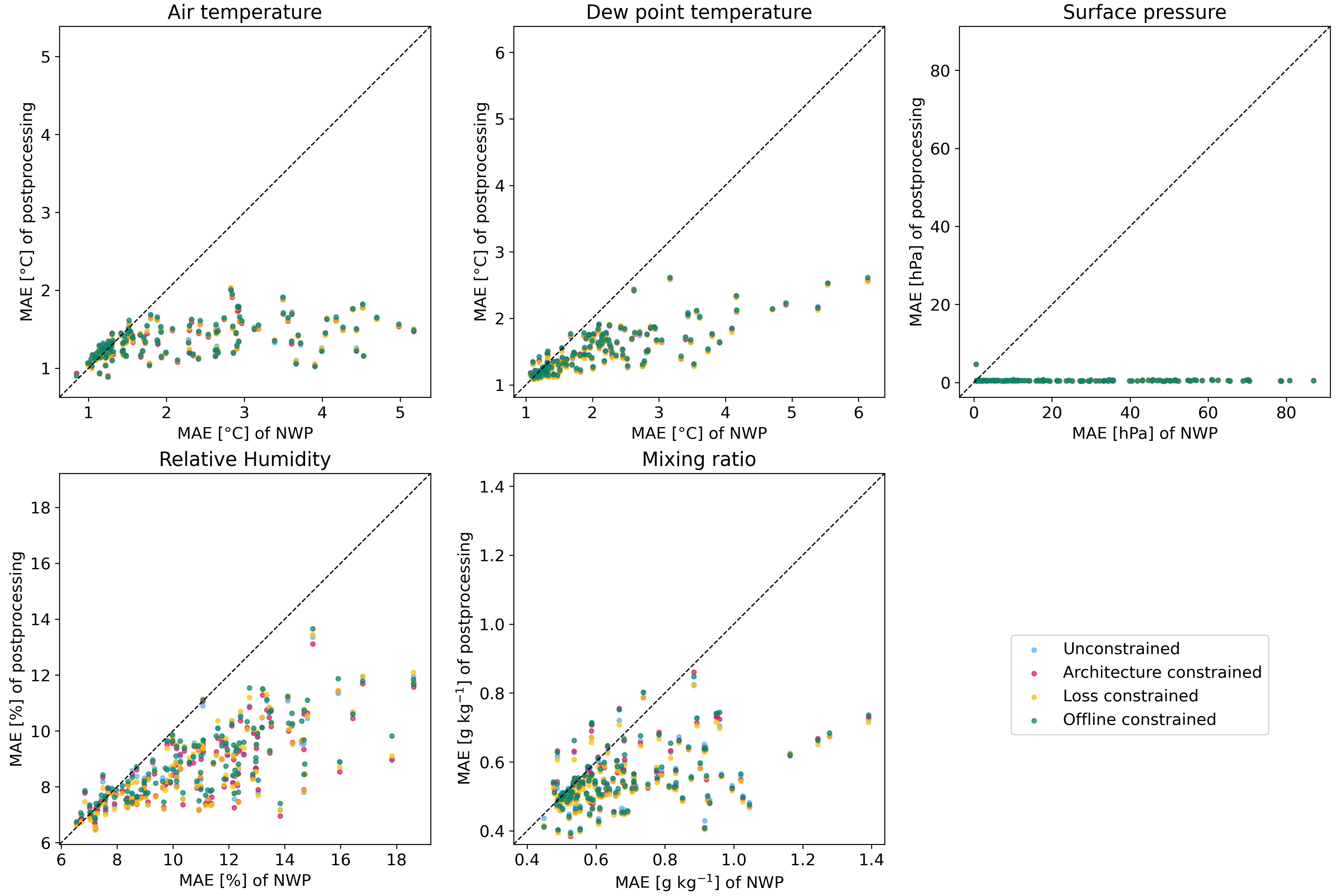}}
    \caption{Scatterplots of the MAE on the test set for the NWP raw forecasts on the x-axis and the postprocessed forecasts on the y-axis.}
\end{figure*}

\begin{figure*}[ht]
    \centerline{\includegraphics[width=\textwidth]{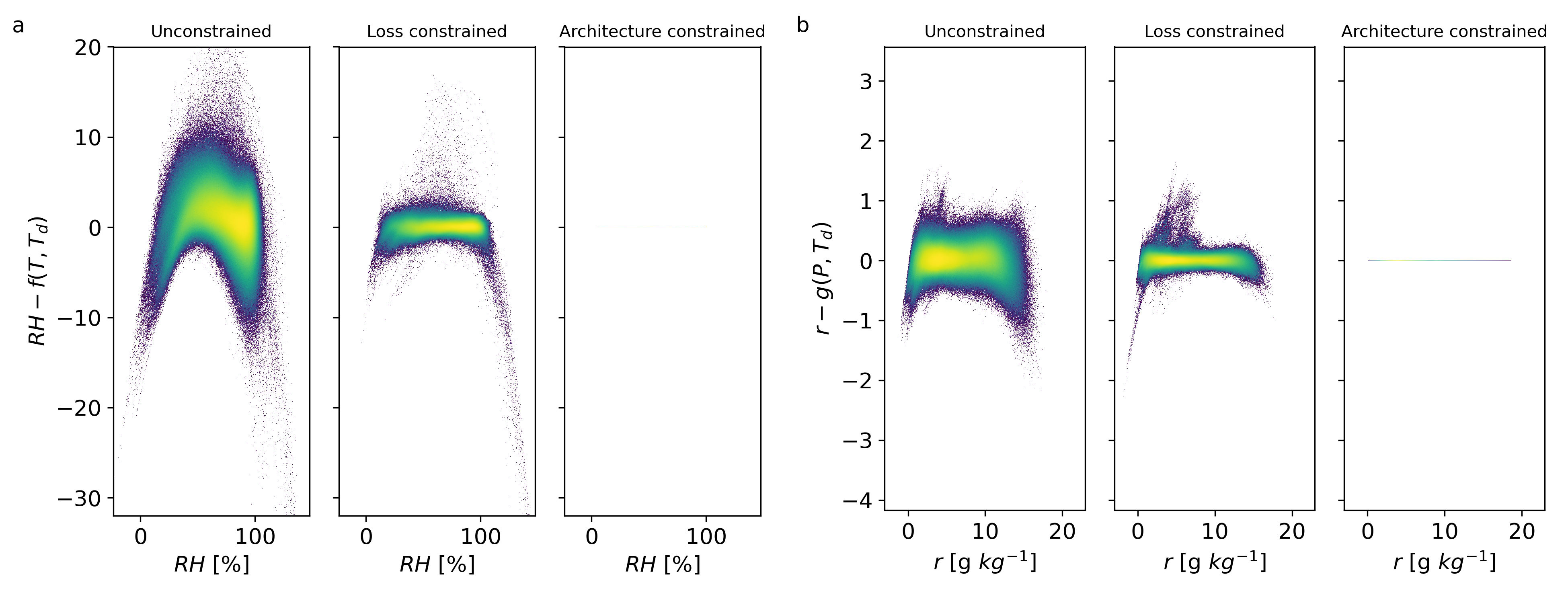}}
    \caption{Scatterplot representing the physical consistency for both relative humidity and water vapor mixing ratio.}
\end{figure*}